\documentclass[prd,preprint,nofootinbib]{revtex4}

\makeatletter
\@ifundefined{lesssim}{\def\lesssim{\mathrel{\mathpalette\vereq<}}}{}
\@ifundefined{gtrsim}{\def\gtrsim{\mathrel{\mathpalette\vereq>}}}{}
\def\vereq#1#2{\lower3pt\vbox{\baselineskip1.5pt \lineskip1.5pt
\ialign{$\m@th#1\hfill##\hfil$\crcr#2\crcr\sim\crcr}}}
\begin{document}

\title{ 
Affleck-Dine baryogenesis and gravitino dark matter
}

\author{Osamu Seto}
\affiliation{
 Department of Physics and Astronomy, University of Sussex, 
 Brighton BN1 9QJ, United Kingdom}

\begin{abstract}
Affleck-Dine baryogenesis in models where the gravitino is both the
 lightest supersymmetric particle and the dark matter candidate is investigated.
For a high enough reheating temperature to produce sufficient gravitinos by
 thermal processes, the observed baryon asymmetry can be explained
 by Affleck-Dine baryogenesis as well as thermal leptogenesis.
On the other hand, if the reheating temperature is not high enough,
 most of the gravitinos must be produced by the decay of the next-to-lightest
 supersymmetric particle (NSP).
Particularly, in the case where Q-balls cannot survive the evaporation,
 the gravitino number density is given by the NSP's thermal relic density.
Interestingly, if Q-balls survive, they can be a source of gravitinos
 via the NSP decay.
Then, we could find a new cosmological interesting region in parameter space
 because the gravitino number density does not relate to
 the NSP's thermal relic density.
\end{abstract}

\maketitle

\section{Introduction}

The origin of dark matter and baryon asymmetry in the Universe is one of prime
 questions in cosmology as well as in particle physics.
The most promising candidate for dark matter is the lightest supersymmetric 
 particle (LSP) owing to the conservation of R parity.
In particular, the lightest neutralino in
 the minimal supersymmetric standard model (MSSM) has been intensively studied.
However, there is another possibility, namely, the LSP is the gravitino
 \cite{EllisGravitino}.
If the reheating temperature after inflation, $T_R$, is as high as $10^{10}$ GeV,
 thermal processes can produce sufficient gravitinos to be the dark matter 
 for the typical gravitino mass $m_{3/2}\simeq 100$ GeV in supergravity models
 \cite{BolzGravitino},
 though the next-to-lightest supersymmetric particle (NSP) decay
 could contribute to the gravitino LSP production as well \cite{FengSuperWIMP}.
Moreover, this is one of solutions to
 the gravitino problem \cite{GravitinoProblem} 
 in supersymmetric models
 \footnote{Other proposals to avoid the gravitino problem involve
 dilution by an entropy production \cite{ThermalInf}, 
 prohibition of the radiative decay of gravitinos \cite{Asaka},
 or the modification of the cosmic expansion 
 in a particular five dimensional model \cite{OS}.}.

The Affleck-Dine (AD) mechanism is a promising source of baryogenesis 
 in supersymmetric models \cite{AffleckDine}, 
 where a condensation of a flat direction which consists 
 of scalar quarks and/or leptons can generate the baryon asymmetry.
In addition, such an Affleck-Dine condensate is typically unstable 
 with respect to spatial perturbations and would 
 form nontopological solitons, Q-balls \cite{QballKusenko, QballEnqvist}. 
When Q-balls decay at a late time
 in a model with gravity mediation of supersymmetry (SUSY) breaking,
 supersymmetric particles are also produced.
This SUSY particle production by Q-ball decay 
 often causes the overproduction of neutralino LSPs 
 unless they annihilate \cite{QballEnqvist, EnqvistMcDonald}.
As a result, available AD fields and the nature of neutralino LSP are 
 considerably restricted.
It is necessary indeed to consider the annihilation of neutralinos 
 \cite{FujiiHamaguchi} 
 or an extension of the MSSM \cite{Singlino}.
 
However, note that this conclusion is obtained under the assumption of
 a neutralino LSP.
Then, it would be natural to consider the gravitino LSP 
 instead of the neutralino to moderate the constraints on AD baryogenesis.
In this paper, we investigate whether AD baryogenesis is compatible with
 gravitino LSP dark matter.

\section{Baryon asymmetry}

The potential of the AD field is, in general, lifted 
 by soft SUSY breaking terms and
 nonrenormalizable terms \cite{Ng, DineRandallThomas}.
The nonrenormalizable superpotential is expressed as
\begin{equation}
W = \frac{\lambda}{nM^{n-3}}\phi^n ,
\end{equation}
where $\lambda$ is the Yukawa coupling and $M$ is some large scale acting 
 as a cut-off.
The total potential of the AD field, including relevant thermal effects, 
 is given by
\begin{eqnarray}
V(\phi) &=& \left(m_{\phi}^2\left[1+K\ln\left(\frac{|\phi|^2}{\Lambda^2}\right)\right] - c_1 H^2 + \sum_{h_i|\phi|<T} h_i^2T^2\right)|\phi|^2 
+\alpha T^4\ln\left(\frac{|\phi|^2}{T^2}\right) \nonumber \\
 && +\left[\left(c_2 H+ Am_{3/2}\right)\lambda\frac{\phi^n}{nM^{n-3}}+ 
 {\rm H.c. } \right] 
 +\lambda^2\frac{|\phi|^{2n-2}}{M^{2n-6}}.
\label{ADPotential}
\end{eqnarray}
Here, $m_{\phi}$ is the soft SUSY breaking scalar mass with 
 radiative correction $K\ln|\phi|^2$, 
 where a flat direction dependent constant, $K$, 
 takes values from $-0.01$ to $-0.1$ \cite{K}.
$\Lambda$ denotes the renormalization scale.
$-c_1 H^2$ represents the negative mass squared 
 induced by the SUSY breaking effect 
 which comes from the energy density of the inflaton, with 
 an order unity coefficient $c_1$ \cite{DineRandallThomas}.
The thermal mass $h_i^2T^2$ can appear when the expectation value of 
 the AD field is relatively small as $h_i|\phi|<T$, 
 where $h_i$ is a coupling of the AD field to other particles \cite{ThermalMass}.
The thermal effect comes from the running of gauge coupling at two loop level,
 and it generates a logarithmic potential, 
 where $|\alpha|$ would be $\mathcal{O}(10^{-2})$ \cite{TwoLoop}.
Terms in the second line in Eq.~(\ref{ADPotential}) contain the A-terms from
 low energy SUSY breaking and those induced by the inflaton respectively.

The baryon number density for the AD field $\phi$ is given by
\begin{equation}
n_b= i q(\dot{\phi}^*\phi-\phi^*\dot{\phi}),
\label{DefADcharge}
\end{equation}
 where $q$ is the baryonic charge for the AD field. 
By using the equation of motion of the AD field,
 Eq.~(\ref{DefADcharge}) can be rewritten as
\begin{equation}
n_b(t) \simeq \frac{1}{a(t)^3}\int^t dt' a(t')^3 \frac{2q m_{3/2}}{M^{n-3}}
{\rm Im} (A \phi^n) ,
\end{equation}
 with $a(t)$ being the scale factor.
When the AD field starts to oscillate around the origin, the angular momentum, 
 which corresponds to the baryon number density through Eq.~(\ref{DefADcharge}), 
 is induced by the relative phase between $A$ and $c_2$.
The generated baryon number density is
\begin{equation}
\left.n_b\right|_{t_{\rm{os}}}
 \simeq 
 \left. \frac{2 q|A| m_{3/2}}{M^{n-3}H}|\phi|^n \sin\delta \right|_{t_{\rm{os}}},
\end{equation}
where $t_{\rm{os}}$ is the time of the start of the oscillation and
 $\sin\delta$ is the effective CP phase. 
Giving $ s = 4\pi^2g_*T^3/90 $, we can read the baryon to entropy ratio as
\begin{eqnarray}
\frac{n_b}{s}
 &=&  \frac{T_R}{4M_P^2H_{\rm{os}}^2} \left.n_b\right|_{t_{\rm{os}}} \nonumber \\
 &\simeq & \frac{q|A|m_{3/2}}{2}\frac{T_R}{M_P^2H_{\rm{os}}^2}
 \frac{|\phi_{\rm{os}}|^n}{M^{n-3}H_{\rm{os}}} \sin\delta ,
\label{b-sRatio}
\end{eqnarray}
 after the reheating time.
Here, $M_P \simeq 2.4 \times 10^{18}$ GeV is the reduced Planck mass.
As one can see, the resultant baryon asymmetry depends on 
 the amplitude of the AD field, $\phi_{\rm{os}}$, as well as 
 the Hubble parameter, $H_{\rm{os}}$, at $t=t_{\rm{os}}$.

In order to evaluate $n_b/s$, we consider the dynamics of the AD field. 
For the moment, we assume that $\alpha$ is positive.
The minimum of the potential during the inflaton dominated stage is given as
\begin{equation}
|\phi| = \left(\sqrt{\frac{c_1}{n-1}}\frac{H M^{n-3}}{\lambda}\right)^{1/(n-2)} 
 \simeq \left(\frac{H M^{n-3}}{\lambda}\right)^{1/(n-2)} ,
\label{PhiInfDom}
\end{equation}
 and the AD field traces the instantaneous minimum after inflation 
 until it becomes unstable.
The AD field begins to oscillate when the effective mass becomes comparable to 
 the Hubble parameter,
\begin{equation}
H_{\rm{os}}^2
 \simeq m_{\phi}^2 + \sum_{h_i|\phi|<T} h_i^2T^2 +\frac{\alpha T^4}{|\phi|^2},
\label{Hos=m}
\end{equation}
 where the effective mass consists of three contributions:
 the soft mass, the thermal mass, and the two loop effect.
Since the temperature of the plasma after inflation behaves as
\begin{equation}
T \simeq T_R\left(\frac{a(t_{\rm{rh}})}{a(t)}\right)^{3/8}
 = T_R \left(\frac{H}{H_{\rm rh}}\right)^{1/4},
\label{Tep:InfDom} 
\end{equation}
 until the reheating is completed,
 with $ 3M_P^2H_{\rm{rh}}^2 = (\pi^2 g_*/30)T_R^4 $ ,
 Eq.~(\ref{Hos=m}) is rewritten as
\begin{equation}
H_{\rm{os}} \simeq
\left\{
\begin{array}{ll}
 & m_{\phi} ,  \\ 
 & \left(\frac{90}{\pi^2 g_*}\right)^{1/6}h_i^{4/3}T_R^{2/3}M_P^{1/3} 
  \quad {\rm if} \quad \left(\frac{\pi^2 g_*}{90}\right)^{1/8}
 \frac{h_i}{T_R^{1/2}M_P^{1/4}}\left(\frac{M^{n-3}}{\lambda}\right)^{1/(n-2)}
 H^{\frac{6-n}{4(n-2)}} < 1 ,  \\ 
 & \left(\frac{90}{\pi^2 g_*}\right)^{(n-2)/2n}
 (\alpha T_R^2 M_P)^{(n-2)/n}(\lambda M^{3-n})^{2/n} ,
\end{array}
\right.
\label{Hos}
\end{equation}
 for the case that the oscillation is induced by the soft SUSY mass,
 the thermal mass, and the two loop effect, respectively.

In the case of $n=4$,
 the promising AD field is the $LH_u$ direction, 
 which has been studied in detail 
 and shown that the sufficient baryon asymmetry can be produced 
 for a wide range of the reheating temperature, including $T_R \sim 10^{10}$ GeV
 \cite{FujiiHamaguchiYanagida}.
Thus, AD baryogenesis via the $LH_u$ direction
 is compatible with gravitino dark matter.
(The issue of NSPs decay will be discussed at the end of this section.)

From now on, we concentrate on the case of $n=6$. 
Then, Eq.~(\ref{Hos}) is rewritten as
\begin{equation}
H_{\rm{os}} \simeq
\left\{
\begin{array}{ll}
& 1 \left(\frac{m_{\phi}}{1 {\rm TeV} }\right) {\rm TeV},  \\ 
& 7 \times 10^3 \left(\frac{h_i}{5\times 10^{-5}}\right)^{4/3}
\left(\frac{T_R}{10^{10} {\rm GeV}}\right)^{2/3} 
\left(\frac{200}{g_*}\right)^{1/6}{\rm TeV} \\
& \qquad {\rm if} \quad
\frac{h_i|\phi|}{T} =
\frac{1}{\lambda^{1/4}} \left(\frac{g_*}{200}\right)^{1/8}
\left(\frac{h_i}{5\times 10^{-5}}\right)
\left(\frac{10^{10} {\rm GeV}}{T_R}\right)^{1/2}\left(\frac{M}{M_P}\right)^{3/4} < 1 ,\\
%
& 3 \times 10^2 \lambda^{1/3}
\left(\frac{\alpha}{10^{-2}}\right)^{2/3}
\left(\frac{T_R}{10^{10} {\rm GeV}}\right)^{4/3}\left(\frac{M_P}{M}\right)
\left(\frac{200}{g_*}\right)^{1/3}{\rm TeV} . \label{HosTwo} 
\end{array}
\right.
\label{Hos:n=6}
\end{equation}
Thus, a thermal mass can be induced for the case with 
 a small coupling $h_i$, a low cut off $M$, or a high reheating temperature,
\begin{equation}
\left(\frac{h_i}{5\times 10^{-5}}\right)\left(\frac{M}{M_P}\right)^{3/4}
 <  \left(\frac{T_R}{10^{10} {\rm GeV}}\right)^{1/2} ,
\label{TRforThermalMass}
\end{equation}
 while
 the two loop effect can be easily important for a high reheating temperature
\begin{equation}
 T_R \gtrsim 10^8 \left(\frac{M}{M_P}\right)^{3/4} {\rm GeV}.
\label{TRforTwoLoopEffect}
\end{equation}
In these last two cases, the AD field undergoes early oscillation.

Using Eqs.~(\ref{PhiInfDom}) and (\ref{Hos:n=6}),
 the baryon asymmetry Eq.~(\ref{b-sRatio}) is estimated as
\begin{eqnarray}
\frac{n_b}{s}
&\simeq& \frac{q|A|\sin\delta}{2\lambda^{3/2}} 
 \frac{m_{3/2}T_R}{m_{\phi}^{3/2}M_P^{1/2}}
 \left(\frac{M}{M_P}\right)^{3/2} \nonumber \\
&\simeq& 2\times 10^{-10} \frac{q|A|\sin\delta}{2\lambda^{3/2}}
 \left(\frac{m_{3/2}}{100\rm{GeV}}\right)
 \left(\frac{10^3 \rm{GeV}}{m_{\phi}}\right)^{3/2}
 \left(\frac{T_R}{100 \rm{GeV}}\right)\left(\frac{M}{M_P}\right)^{3/2} ,
\label{BaryonAsymNoEarly}
\end{eqnarray}
for the case of $H_{\rm{os}} \simeq m_{\phi}$,
\begin{eqnarray}
\frac{n_b}{s}
 &\simeq& \frac{q|A|\sin\delta}{2\lambda^{3/2}}
 \left(\frac{\pi^2 g_*}{90}\right)^{1/4}\frac{m_{3/2}}{f_i^2M_P}
 \left(\frac{M}{M_P}\right)^{3/2} \nonumber\\
 &\simeq& 9 \times 10^{-11}\frac{q|A|\sin\delta}{2\lambda^{3/2}}
 \left(\frac{m_{3/2}}{100 {\rm GeV}}\right)
 \left(\frac{10^{-9/2}}{h_i}\right)^2\left(\frac{g_*}{200}\right)^{1/4}
 \left(\frac{M}{10^{-2}M_P}\right)^{3/2} ,
\label{BaryonAsymThermalMass}
\end{eqnarray}
for the case that the oscillation is driven by the thermal mass,
\begin{eqnarray}
\frac{n_b}{s}
&\simeq&
 \frac{q|A|\sin\delta}{2\lambda^2}\left(\frac{\pi^2 g_*}{90}\right)^{1/2}
 \frac{m_{3/2}}{\alpha T_R}\left(\frac{M}{M_P}\right)^3 \nonumber\\
&\simeq& 0.5\times 10^{-11}
 \frac{q|A|\sin\delta}{2\lambda^2}\left(\frac{g_*}{200}\right)^{1/2}
 \left(\frac{10^{10}\rm{GeV}}{T_R}\right)
 \left(\frac{M}{10^{-2} M_P}\right)^3
 \left(\frac{m_{3/2}}{10^2\rm{GeV}}\right)\left(\frac{10^{-2}}{\alpha}\right),
\label{BaryonAsymTwoLoopThermal}
\end{eqnarray}
for the case that the oscillation is driven by the two loop effect.
For the cases that the AD field undergoes early oscillation
 due to thermal effects, we normalized the reheating temperature as 
 $T_R \simeq 10^{10}$ GeV to produce gravitinos appropriately and, then, 
 found that the cut-off scale must be suppressed as $M \lesssim 10^{-2}M_P$ 
 in order to satisfy the conditions Eqs.~(\ref{TRforThermalMass}) 
 and (\ref{TRforTwoLoopEffect}) properly.
On the other hand, in Eq.~(\ref{BaryonAsymNoEarly}), obviously
 the reheating temperature is not high enough to
 produce gravitinos by thermal processes.
Hence, if this is the case in the model with gravitino LSP, 
 most of the gravitino dark matter must be produced by NSPs decay.
 
Next, consider the dynamics of the AD field after the oscillation for the case
 that the AD field undergoes early oscillation due to thermal effects.
While the AD field oscillates due to the two loop thermal effect,
 its amplitude decreases \cite{FujiiHamaguchiYanagida, DolgovKohriSetoYokoyama}
\begin{eqnarray}
 |\phi| &\simeq& |\phi_{\rm{os}}|\left(\frac{a_{\rm{os}}}{a(t)}\right)^{9/4} \\
 &=& |\phi_{\rm{os}}|\left(\frac{H}{H_{\rm{os}}}\right)^{3/2}
 \qquad {\rm for \quad matter \, (inflaton)\,\,  dominated}.
\end{eqnarray}
Using $ |\phi_{\rm{os}}| \simeq (H_{\rm{os}}M^3/\lambda)^{1/4} $, we find
\begin{eqnarray}
 \frac{h_i|\phi|}{T} &=& \frac{|\phi_{\rm{os}}|}{T_R}
 \left(\frac{a_{\rm{os}}}{a(t)}\right)^{9/4}\left(\frac{a(t)}{a_{\rm{rh}}}\right)^{3/8} \nonumber \\
 &\simeq& \frac{h_i(M^3H_{\rm{rh}})^{1/4}}{\lambda^{1/4}T_R}
 \left(\frac{H}{H_{\rm{os}}}\right)^{5/4}
 \equiv
 \left.\frac{h_i|\phi|}{T}\right|_{t_{\rm{os}}}\left(\frac{H}{H_{\rm{os}}}\right)^{5/4}.
\end{eqnarray}
If $\left.h_i|\phi|/T\right|_{t_{\rm{os}}}< 1$ the AD field starts to oscillate 
 due to the thermal mass, 
 whereas if $\left.h_i|\phi|/T\right|_{t_{\rm{os}}}> 1$ 
 the AD field starts to oscillate due to the two loop effect. 
In the latter case, the thermal mass term appears soon after
 the AD field starts to oscillate by the two loop thermal effect, because
 $h_i|\phi|/T$ decreases after the AD field began to oscillate.
After the Hubble parameter becomes
\begin{equation}
H \simeq 
 \left\{ 
 \begin{array}{ll}
 \left(\frac{5\times 10^{-5}}{h_i}\right)^{4/5}
 \left(\frac{M_P}{M}\right)^{3/5} H_{\rm{os}} & {\rm for}
  \quad h_i > 5 \times 10^{-5}, \\
 \left(\frac{10^{-3}}{h_i}\right)^{4/5}
 \left(\frac{10^{-2}M_P}{M}\right)^{3/5} H_{\rm{os}} & {\rm for}
  \quad h_i > 10^{-3}, \\
\end{array} 
 \right.
\end{equation}
the thermal mass is induced and the AD field oscillates due to the thermal mass.

Hence, for both cases,
 since the AD field oscillates around the origin due to the thermal mass term,
 the AD field could evaporate before Q-balls are formed.
The evaporation of an AD field has been studied by many authors, for example, 
 \cite{ThermalMass, TwoLoop, Yokoyama, Dolgov}.
In fact, for instance, adopting the scattering rate $\Gamma \sim h^4 T$, 
 we find
\begin{equation}
\left.\frac{\Gamma}{H}\right|_{t_{\rm{rh}}} \simeq 1
 \left(\frac{h}{10^{-2}}\right)^4\left(\frac{10^{10} {\rm GeV}}{T_R}\right) ,
\end{equation}
 and the AD condensate can evaporate when the reheating finishes.
For $T_R \simeq 10^{10}$ GeV,
 $H_{{\rm rh}}$ is given as $H_{{\rm rh}} \simeq 10^2$ GeV and,
 then, the AD field oscillates due to the thermal mass
 $10^8 {\rm GeV}(h_i/10^{-2})(T_R/10^{10} {\rm GeV})$, which is much
 larger than the soft mass.
Thus, the AD field would evaporate before the spatial instability
 which leads to Q-balls is induced by the soft mass,
 in the case that the AD field oscillates due to thermal effects.
However, for a very small coupling $h_i$, the evaporation of AD field would
 be delayed and Q-balls might be formed.
Even if Q-balls are formed, as we will see,
 they would evaporate soon
 because the charge is small due to the small value of $M$.

Therefore, the only requirement for 
 the cases where the AD field undergoes early oscillations
 is that the thermal relic abundance of the NSPs
 should not be in conflict with constraints from their late decay,
 just as in thermal leptogenesis \cite{FukugitaYanagida}.
Thus, these scenarios can be consistent
 because a part of the parameter space in a model with a gravitino LSP remains
 consistent \cite{EllisGravitinoDM, Roszkowski, New}.

On the other hand, in the case of $H_{\rm{os}}\simeq m_{\phi}$, 
 as is well known, 
 a relatively low reheating temperature is required to explain 
 the appropriate baryon asymmetry.
In this case, Q-balls are formed for most of the flat directions.
In addition, most of the gravitinos must be generated non-thermally 
 via NSP decay,
 because the thermal production of gravitinos is not enough
 to explain dark matter for such a low reheating temperature.
In the next section, we will discuss the issue of Q-balls and NSPs.

\section{Q-ball decay and NSP abundance}

Here, we briefly summarize important properties of Q-balls.
The radius of a Q-ball, $R$, is estimated as $R^2 \simeq 2/(|K|m_{\phi}^2)$
 \cite{QballEnqvist}.
The charge is roughly estimated as
\begin{eqnarray}
Q \simeq \frac{4}{3}\pi R^3 n_{b}(t_i)
 \simeq \frac{4}{3}\pi R^3 \left(\frac{H_i}{H_{\rm{os}}}\right)^2 
 n_{b}|_{t_{\rm{os}}} ,
\end{eqnarray}
 where the suffix $i$ represents the time when the spatial imhomogeneity
 becomes nonlinear.
According to numerical calculations, the Q-ball charge is expressed as
\begin{equation}
Q \simeq \bar{\beta}\left(\frac{|\phi_{\rm{os}}|}{m_{\phi}}\right)^2 \times 
\left\{
\begin{array}{ll}
\epsilon  \;\quad {\rm for} \quad \epsilon \gtrsim \epsilon_c \\
\epsilon_c  \quad {\rm for} \quad \epsilon < \epsilon_c 
\end{array}
\quad ,
\right.
\end{equation}
where $ \epsilon_c \simeq 10^{-2} $, $\bar{\beta}= 6\times 10^{-3}$ and
$ \epsilon \equiv n_b/n_{\phi}|_{t_{\rm{os}}} \simeq
(2q|A|/\lambda)(m_{3/2}/m_{\phi})\sin\delta $ \cite{KasuyaKawasaki}.
Hence, the Q-ball charge can be evaluated as
\begin{eqnarray}
Q &\sim& 6\times 10^{-3} \frac{2q|A|\sin\delta}{\lambda^{3/2}}
\frac{m_{3/2}M_P^{3/2}}{m_{\phi}^{5/2}}
\left(\frac{M}{M_P}\right)^{3/2} \nonumber \\
 &\simeq& 1 \times 10^{20} \frac{q|A|\sin\delta}{\lambda^{3/2}}
\left(\frac{m_{3/2}}{100 {\rm GeV}}\right)
\left(\frac{1 {\rm TeV}}{m_{\phi}}\right)^{5/2}\left(\frac{M}{M_P}\right)^{3/2}.
\label{QballCharge}
\end{eqnarray}
A part of the charge of a Q-ball can be evaporated 
 by the interaction with particles in the thermal bath.
The evaporated charge is estimated as $\Delta Q =\mathcal{O}(10^{18})$
 \cite{BanerjeeJedamzik}.
Thus, if the charge of a Q-ball is larger than $\mathcal{O}(10^{18})$,
 Q-balls can survive the evaporation.
For a Q-ball charge of $M \simeq M_P$ in Eq.~(\ref{QballCharge}), 
 the decay temperature of Q-ball is $T_d \simeq 1$ GeV \cite{Td}, 
 which is lower than the typical freeze out temperature of WIMPs, $T_f$.
NSPs are then generated by the Q-ball decay.

The NSP abundance after Q-balls decay (until NSP decay) is given as
\begin{equation}
Y_N(T) \equiv \frac{n_{NSP}}{s} \simeq \left[ \frac{1}{Y(T_d)}
+\sqrt{\frac{8\pi^2 g_*(T_d)}{45}}\langle\sigma v\rangle M_P(T_d-T) \right]^{-1},
\end{equation}
 by solving the relevant Boltzmann equation \cite{FujiiHamaguchi}, 
\begin{equation}
\dot{n_N}+3Hn_N=-\langle \sigma v\rangle n_N^2,
\end{equation}
 where the NSP would be the lightest neutralino or stau. 
In the case that the annihilation is inefficient
 after NSPs are produced by Q-ball decay, the final abundance is given as
\begin{eqnarray}
Y_N \simeq Y(T_d) =
 3\left(\frac{N}{3}\right)f_B\frac{n_b}{s}
 = 3\times 10^{-10}
 \left(\frac{N}{3}\right)\left(\frac{f_B}{1}\right)
 \left(\frac{n_b/s}{10^{-10}}\right) ,
\label{YN:NoAnnihilation}
\end{eqnarray}
 because at least three supersymmetric particles per one baryonic charge 
 are produced.
Here, $f_B$ is a ratio of baryon number in Q-balls to the total baryon number.
Recalling 
 $ \rho_c/s \simeq 1.7 \times 10^{-9}(h/0.7)^2 $ GeV, or equivalently,
\begin{equation}
\frac{\rho_{DM}}{s}
 \simeq 3.9 \times 10^{-10}\left(\frac{\Omega_{DM}}{0.23}\right)
\left(\frac{h}{0.7}\right)^2 \rm{GeV},
\end{equation}
 Eq.~(\ref{YN:NoAnnihilation}) leads to the well known result that
 the mass of the dark matter particle must be about $1$ GeV 
 for $n=6$ Affleck-Dine baryogenesis. 
Thus, the possibility that the Affleck-Dine baryogenesis accounts for 
 both the dark matter and the baryon asymmetry due to Q-ball decay 
 was given up for the neutralino in the MSSM \cite{QballEnqvist}.
For the gravitino, although the possibility of $m_{3/2}\simeq 1$ GeV
 is not excluded experimentally, it looks somewhat unnatural and unlikely 
 in the gravity mediated SUSY breaking.

In the case that the annihilation occurs after the NSPs are produced,
 the final abundance of NSPs for $T \ll T_d$ is given as
\begin{equation}
Y_N \simeq
\left[\sqrt{\frac{8\pi^2 g_*(T_d)}{45}}\langle\sigma v\rangle M_PT_d \right]^{-1}.
\end{equation}
In this scenario, all of the dark matter gravitinos must be produced from NSPs
 when they decay.
By imposing the constraint on $Y_N$ 
 from $\rho_{NSP} \simeq \rho_{LSP} = \rho_{DM}$, we obtain
\begin{equation}
Y_N \simeq 3 \times 10^{-12}
\left(\frac{3\times 10^{-8} {\rm GeV}^2}{\langle\sigma v\rangle}\right)
\left(\frac{1 {\rm GeV}}{T_d}\right)\left(\frac{10}{g_*(T_d)}\right)^{1/2}
 \sim 4 \times 10^{-12}  \left(\frac{100 {\rm GeV}}{m_{DM}}\right),
\end{equation}
that is, a large annihilation cross section
 $\langle\sigma v\rangle\simeq 10^{-8} - 10^{-7} {\rm GeV}^2$ is required
 for $T_d \sim 1$ GeV.
Here, $m_{DM}$ is the mass of the dark matter particle.
In order to realize such a large annihilation cross section,
 neutralino NSP should not be Bino-like but Higgsino-like \cite{BolzGravitino}.
Similarly, in the stau NSP case,
 the stau can have such large annihilation cross section 
 \cite{EllisCoannihilation}.
In addition, the late decay of NSPs must not disturb 
 the Big Bang Nucleosynthesis (BBN) and 
 Cosmic Microwave Background radiation (CMB).
Though this is a very severe constraint, a small region in the parameter space
 still looks to be consistent, according to recent studies 
 \cite{FengSuperWIMP, Roszkowski}.
However, one should notice that these results
 are not simply applicable to this scenario
 because the relic abundance of the NSP is thermally determined there.
A detailed study on this issue, for this particular case, is necessary.

\section{The case of two loop induced potential with a negative coefficient}

Finally, we consider the case that $\alpha$ in the two loop effect 
 in Eq. (\ref{ADPotential}) is negative
 \footnote{This case in the gauge mediated SUSY breaking has been studied
 in Ref. \cite{Neg}.}.
After inflation the AD field traces the instantaneous minimum, 
 Eq.~(\ref{PhiInfDom}), as in previous cases.
However, after a while, if Eq.~(\ref{TRforTwoLoopEffect}) is satisfied
 \footnote{If not so, 
 it becomes the case of $H_{os} \simeq m_{\phi}$ already mentioned 
 in the previous sections.}, 
 the two loop effect exceeds the mass of order of $H$.
After that, the instantaneous minimum is given by 
\begin{equation}
(-\alpha)\frac{T^4}{|\phi|^{2}}
 \simeq \frac{(n-1)\lambda^2|\phi|^{2n-4}}{M^{2n-6}} \,\, .
\label{NewMin}
\end{equation}
While the AD field is trapped by the two loop effect, 
 $h_i|\phi|/T$ increases and is always larger than unity.
Hence, the thermal mass never appears in this case (See, section II).
When $(-\alpha)T^4/|\phi|^{2}$ becomes comparable to $m_{\phi}^2$,
 the AD field starts to oscillate.
The temperature of the plasma and the Hubble parameter at that time are given as
\begin{eqnarray}
T_{os} &\simeq& \left(\frac{ m_{\phi}^{2\frac{n-1}{n-2}}M^{2\frac{n-3}{n-2}}}
{(-\alpha)[(n-1)\lambda^2]^{\frac{1}{n-2}}}\right)^{1/4} , \\
H_{os} &\simeq& \sqrt{\frac{\pi^2 g_*}{90}}
 \frac{m_{\phi}^{2\frac{n-1}{n-2}}M^{2\frac{n-3}{n-2}}}
 {(-\alpha)[(n-1)\lambda^2]^{\frac{1}{n-2}}T_R^2M_P} ,
\end{eqnarray}
from Eqs.~(\ref{Tep:InfDom}) and (\ref{NewMin}).
Thus, the resultant baryon asymmetry is estimated as
\begin{eqnarray}
\frac{n_b}{s} &\simeq& \frac{q|A|\sin\delta}{2}
 [(n-1)\lambda^2]^{-\frac{n-6}{2n-4}}
 \left(\frac{\pi^2 g_*}{90}\right)^{-3/2}
 \frac{(-\alpha)^3m_{3/2}M_PT_R^7}{m_{\phi}^{\frac{5n-6}{n-2}}
 M^{4\frac{n-3}{n-2}}} \\
 &\simeq& 10^{-10}\frac{q|A|\sin\delta}{2}\left(\frac{g_*}{200}\right)^{-3/2}
 \left(\frac{-\alpha}{10^{-2}}\right)^3
 \left(\frac{m_{3/2}}{10^2 {\rm GeV}}\right)
 \left(\frac{1 {\rm TeV}}{m_{\phi}}\right)^6
 \left(\frac{10^{14} {\rm GeV}}{M}\right)^3
 \left(\frac{T_R}{10^{5.4} {\rm GeV}}\right)^7   
 \nonumber 
\end{eqnarray}
for $n=6$.
Here, one may notice that this result is very sensitive
 to the variation of parameters such as $m_{\phi}$ and $T_R$.
We found that the cut-off of the nonrenormalizable term has to be reduced to
 $M < 10^{15}$ GeV, which can be rewritten as $T_R < 10^6$ GeV also, 
 to satisfy Eq.~(\ref{TRforTwoLoopEffect}) while keeping $n_b/s \simeq 10^{-10}$.
Formed Q-balls cannot survive the evaporation 
 because of the small charge by the reduced $M$.
Hence, Q-ball formation does not affect the viability of AD baryogenesis
 if the baryon asymmetry is produced appropriately.
On the other hand, in this case also,
 most of gravitinos must be produced by the NSP decay.
Here, notice that the NSP density is determined by the thermal relic density 
 unlike that in the previous case 
 where NSPs are produced non-thermally by Q-ball decay.
Therefore, the gravitino dark matter in this case is nothing but 
 the ``SuperWIMP scenario'' proposed by Feng et al \cite{FengSuperWIMP}.
In other words, 
 Affleck-Dine baryogenesis 
 with a low reheating temperature and evaporated Q-balls, in this case, is 
 compatible with the SuperWIMP scenario.
The constraints derived in previous works
 \cite{FengSuperWIMP, EllisGravitinoDM, Roszkowski}
 are directly applicable to this case.
It is found that this scenario is not viable 
 at least within the Constraint MSSM, 
 according to the latest work \cite{New}.

\section{Summary}

We have studied Affleck-Dine baryogenesis in a model with gravitino LSP.
The estimation of baryon asymmetry depends on whether the AD field undergoes 
 early oscillation by thermal effects for a high reheating temperature.

A high reheating temperature is acceptable and preferable 
 if we assume that gravitinos are the LSP.
We have shown that the appropriate baryon asymmetry can be obtained, 
 because the baryon asymmetry is reduced with the help of early oscillation.
Hence, we point out that not only thermal leptogenesis \cite{FukugitaYanagida}
 but also Affleck-Dine baryogenesis is viable and compatible with 
 the gravitino dark matter.

In the cases that the AD field begins to oscillate by the soft mass,
 namely the reheating temperature is relatively low,
 while the baryon asymmetry can be explained, 
 most of the gravitinos must be produced by NSP decay.
Then, the important topic is the NSP decay and 
 the conclusions depend on the nature of the NSP and 
 the constraints from BBN and CMB.
If the NSP's annihilation cross section is small, 
 we are again faced with the overproduction of LSPs
 as in the case of neutralino LSP, unless we accept $m_{3/2}\simeq 1$ GeV 
 besides avoiding the constraints from NSP decay.
For the cases when NSPs annihilate,
 if Q-balls can not survive the evaporation, 
 gravitinos must be produced by the decay of NSPs 
 whose abundance is given by the thermal relic density.
If Q-balls can survive the evaporation, NSPs are decay products of Q-balls.
Hence, such long-lived Q-balls could provide a new possibility for
 gravitino dark matter from NSP decay,
 and enlarge the available region of parameter space.
Needless to say, that detailed studies on BBN and CMB constraints 
 for each specific model are
 one of the most important future projects. 

%
\section*{Acknowledgments}
We would like to thank John McDonald for fruitful discussions
 and Leszek Roszkowski for kind correspondence.
We are grateful to Beatriz de Carlos for reading the manuscript.
This work is supported by PPARC. 



\end{document}